\documentclass[aps,prl,twocolumn,groupedaddress,showpacs]{revtex4}
\usepackage{dcolumn}
\usepackage{epsfig}
\usepackage{amsmath}

\begin{document}

\title{
Reliable First-Principles Alloy Thermodynamics via Truncated Cluster Expansions
}

\author{Nikolai A. Zarkevich}
\email{zarkevic@uiuc.edu}
\author{D.D. Johnson}
\affiliation{Departments of Physics and Materials Science \& Engineering, and
Frederich Seitz Materials Research Laboratory,
University of Illinois, Urbana-Champaign, IL 61801}

\date{April 4, 2003} 

\begin{abstract}
In alloys cluster expansions (CE) are increasingly used to combine
first-principles electronic-structure calculations and Monte Carlo methods 
to predict thermodynamic properties.
As a basis-set expansion in terms of lattice geometrical clusters 
and effective cluster interactions, the CE is exact if infinite, 
but is tractable only if truncated.
Yet until now a truncation procedure was not well-defined and
did not guarantee a reliable truncated CE.
We present an optimal truncation procedure for CE basis sets that 
provides reliable thermodynamics.
We then exemplify its importance in Ni$_3$V, where
the CE has failed unpredictably, and now
show agreement to a range of measured values, 
predict new low-energy structures, 
and explain the cause of previous failures.
\end{abstract}
\pacs{05.10.-a,64.70.Kb,81.30.Bx}

\maketitle

{\par} 
The cluster expansion (CE) is increasingly used as a valuable tool
for predicting and interpreting thermodynamic effects in a wide class 
of materials and problems, including
precipitation \cite{PRL88p125503,prl86p5518,prb-alag,Ag2Al},
solubility limits \cite{prl86p448}, ionic diffusion \cite{prb64p184307},
surface alloying \cite{prl87p236102} and patterning \cite{prl88p186101},
vacancy \cite{prl87p275508} and chemical \cite{prb-alag,prl87p165502} 
ordering.
As a means for multiscaling based on  density-functional theory (DFT)
electronic-structure energetics, the CE is a basis-set expansion
in $n$-body clusters (associated with $n$ Bravais lattice points)
and effective cluster interactions (ECI) that specify
configurational energies.
Except for implicit DFT errors in the energy database, the CE is exact 
for an infinite basis,  but impractical if not vastly truncated 
\cite{ClusterExpansion84,PRB48p14013y1993}.
Although there are many successes, a truncated CE can and has
unpredictably failed.

{\par} We present a new method for an optimal truncation of the basis set that 
gives reliable thermodynamics.
We then detail its importance in face-centered-cubic (fcc) Ni$_3$V, 
a system with order-disorder  
transition from disordered $A1$ phase to  ordered DO$_{22}$ phase at 
$T_{c}$ of $1318\,$K \cite{handbookPDbinaryAlloys2000}.
Previous CE for Ni$_3$V
\cite{PRB49p16058,PRB50p6626,PRB52p8813} had errors 
of $40-1000\%$ for a range of thermodynamic properties,
prompting a search for missing physics \cite{PRB52p8813}.
We show that our new method allows 
more reliable predictions, including that of key low-energy 
configurational excitations.
As a synopsis, we compare in Table~\ref{tExptCE} our CE results,
along with the previous ones,
with experimental values of $T_{c}$ and $\Delta E^{L1_{2}-DO_{22}}_{SRO}$, 
the energy difference between DO$_{22}$ and metastable L1$_{2}$ as 
assessed from the short-range order (SRO) measurements
\cite{Finel1994,PRB50p12980}.
The new CE now agrees with a 
range of experimentally assessed values (more below).
We find that prior failure in Ni$_3$V is due to inappropriate truncation of 
the cluster basis set and overfitting to get the ECI -- underscoring
again the need for careful application of basis-set methods.
We have tested this new CE method on a few
cubic and non-cubic binaries and ternaries and found it to be 
especially important when multibody ECI are significant.

\begin{table}[b]
\caption{\label{tExptCE}
New truncated CE (CE$_2$ and CE$_3$) and 
experimental \cite{Finel1994} values of $T_c$ (Kelvin) and 
the $\Delta E_{SRO}^{L1_2 - DO_{22}}$  (meV/atom) assessed from SRO,
along with the former CE \cite{PRB49p16058} and 
CPA \cite{Johnson2000} results. Details in text.
}
\begin{tabular}{l|cc|l|c|c}
\hline
& CE$_2$ & CE$_{3}$ & Expt. 
& old CE 
& CPA 
\\
\hline
$T_c$ (K) & 1335 & 1370 & 1318 & 1900 & \\
$\Delta E_{SRO}^{L1_2 - DO_{22}}$ & 22$\pm16$ & 17$\pm15$ 
			& 12$\pm 5$ & 101 & 7-12 \\
\hline
\end{tabular}
\end{table}

\begin{figure}[t]
 \includegraphics[width=67mm]{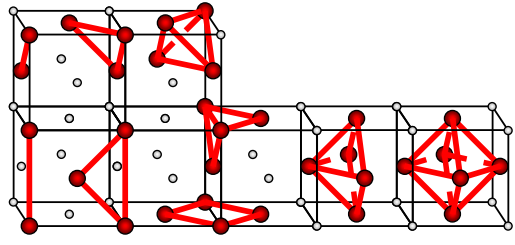} 
 \caption{\label{FigClusters}
The 2- to 6-body clusters for $1^{st}$ and $2^{nd}$ fcc neighbors.
Including clusters up to 4-body in upper set (6-body in both sets) 
form a locally complete basis in the range of $1^{st}$ ($2^{nd}$) neighbors.  
These clusters form tetrahedron (tetrahedron-octahedron)
approximations used in cluster-variation method \cite{ClusterExpansion84}.
 }
\end{figure}

{\par} {\em Cluster Expansions:}
Any alloy configuration may be
represented by a set of occupation variables
$\{\xi_p^{\alpha}\}$, with $\xi_p^{\alpha} = 1 (0)$ if the site $p$
{\em is (is not)} occupied by an $\alpha$-atom.
Composition $c^{\alpha}$ is the thermal- and site-average of
$\{\xi_p^{\alpha}\}$ with $0 \le c^{\alpha} \le 1$.
The energy of any atomic configuration $\sigma $
can be written in a CE \cite{ClusterExpansion84}
using the $n$-body ECI $V^{(n)}_{f}$:
\begin{equation}
\label{H1}
E_{CE} (\sigma ) = V^{(0)} + \sum_{n,f,d} V^{(n)}_{f} \bar{\Phi}^{(n)}_{fd} (\sigma ) .
\end{equation}
Sums are over symmetry--{\em distinct} $(n,f)$ and {\em equivalent} 
$(d=1,...,D_f^{(n)}$, the degeneracy) clusters.
A CE basis can be also presented as a product of orthonormal 
Chebychev polynomials 
based on $\xi_p^{\alpha}$ \cite{ClusterExpansion84}. 
The $n$-site correlation function
$ \bar{\Phi}^{(n)}_{fd}  =
\langle \xi_{p'} \xi_{p''} \ldots \xi_{p^{(n)}} \rangle $
is given by an ensemble average over the fixed sets
$ \{ p \}^{(n)}_{fd} $ defining the $n$-body clusters of type $(f,d)$,
see Fig.~\ref{FigClusters}. 
When evaluated above  $T_{c}$, $\bar{\Phi}^{(2)}$, for example, are 
related to the SRO.
If the ECI are known, then the energy of any
configuration can be predicted. 

{\par} A CE can be truncated if there is rapid convergence of the ECI 
$V^{(n)}_f$ with increasing distance $r$ ({\it e.g.}, 
as measured by cluster radius of gyration or circumscribed sphere)
and with increasing number of sites $n$ in a cluster $f$, 
{\it i.e.} smaller $n$-body clusters are more physically important.
Also
$V^{(n)}_f$ for $n$$>$$n_0$ uncorrelated sites have their 
contributions to  (\ref{H1}) suppressed by 
$\bar{\Phi}^{(n)}$$ \sim$$ c^{n}$, {\emph i.e.} $V^{(n)}\Phi^{(n)}$$ \to$$ 0$,
and can be neglected.
The magnitudes of $V^{(n)}$ typically become smaller 
for larger $n$, although for some systems ECI convergence is not rapid: such is
 Li$_{x}$NiO$_{2}$ where Jahn-Teller distortions control Li-vacancy 
ordering and ionic conduction and are reflected only in 
long-range multibody ECI \cite{PRB63pg144107}.
%
For a truncated CE,  ECI are obtained via 
{\em structural inversion} \cite{SIM-deFontain,SIM-Johnson} 
at fixed $c$ for $c$-dependent (canonical) ECI
or versus $c$ for $c$-independent (grand-canonical) ECI;
these sets of ECI are related \cite{ECI1,ECI2}.
First, a set of $N$ fully-ordered (few atoms per cell) structures
is {\em somehow} chosen and their DFT energies 
$E^i_{DFT}$ ($i=1..N$) are calculated.
Then, a set of $M$ clusters ($M$$<$$N$) is {\em somehow} picked
for use in (\ref{H1}) and $\bar{\Phi}$ are calculated for each structure.
A system of $N$ linear equations (\ref{H1})
with $M$ unknown ECI is solved by least-squares (LS) fitting -- 
which unavoidably includes DFT errors  in energy differences.
As is obvious, the  sets of structures and clusters used to get 
the ECI are not uniquely defined. 

{\par} \emph{New Method:} 
Here we propose a method that, given a set of structural energies,
unambiguously defines a set of clusters (and ECI) to provide an 
 optimal truncated CE and yield reliable thermodynamics.
First, we note that
if $V^{(n)}(r$$>$$r^{(n)}_0) \equiv 0$, 
the truncated CE basis with local compact support that 
includes  \emph{all} clusters in $r^{(n)}_0$ is \emph{locally complete} 
and exact; whereas, if $V^{(n)}(r$$>$$r^{(n)}_0) \approx 0$, this truncated CE is 
approximate and has an error.
With no {\em a priori} knowledge of which clusters are required 
to represent well a given alloy, the CE error is minimal, in a 
Rayleigh-Ritz variational sense, if all admissible $n$-body clusters
(basis functions) of smaller spatial extent ($r$$\le $$r^{(n)}_0$) are 
included before the larger ones.
In brief, having $N$ $E^i_{DFT}$ to be fitted, we can establish
a variational CE basis by simple rules  {\em ad vitam aut culpam}
that implement easily computationally: 
\begin{enumerate}
  \item \vspace{0mm}
	If an $n$-body cluster is included, then include
	{\em all $n$-body clusters of smaller spatial extent}.
  \item \vspace{-1mm}
	If a cluster is included, include {\em all its subclusters}.
  \item \vspace{-1mm}
	To prevent both under- and over-fitting,
	{\em minimize} 
	\cite{PRB53p6137CV,JPhaseEq23pg384} the 
	cross-validation (CV) error \cite{CVscore1,CVscore2}:
	\vspace{-1mm}
\begin{equation}
 \label{CV}
  CV^2 = \frac{1}{N} \sum_{i=1}^{N} ( E^i_{DFT} - E^{(i)}_{CE'} )^2 .
\end{equation}
\end{enumerate}\vspace{-1mm}
$E^{(i)}_{CE'}$ in (\ref{CV}) is {\em predicted}  by a fit to $N$$-$$1$ DFT energies
{\em excluding} $E^i_{DFT}$, rather than to all $N$ as in a LS fit.
(This is an ``exclude 1'' CV,  whereas an ``exclude 0'' CV is a LS fit.)
While LS measures the error in reproducing known values of $E_{DFT}^i$, 
CV error estimates an uncertainty of {\em predicted} values. 
Both too few (underfitting) or too many 
(overfitting) parameters give poor prediction.
The new Rule 1, with well-established Rule 2, now makes it easy to define 
uniquely all clusters in a  
truncated, variational CE basis by the number of $n$-body clusters 
(or the size of the largest $n$-body cluster) for each $n\le n_0$. 
In particular, Rules 1 and 2 permit a hierarchy of ranges for $n$-body clusters, 
i.e., $r^{(n)}_0$$\ge$$r^{(n+1)}_0$ for all $n$, giving a locally  complete basis 
for strict equalities, while the inequality 
({\emph e.g.}, more 2-bodies, fewer 3-bodies, even fewer 4-bodies, etc.)
allows for fewer clusters.
Because shorter-ranged and lower-order ECI are more important,
Rule 1 [Rule 2] prohibits excluding more important ECI and transferring 
their weight to less important longer-ranged [higher-order] ones.
{\par} Once constructed, an optimal CE can be used to predict energy of any 
structure  within the CV error. 
The CE is valid if the lowest structural energies (including the ground-state)
and fully-disordered state energies are correct within the accuracy given by 
the CV error. A valid optimal CE provides reliable thermodynamics.

\begin{table}[b]
\caption{\label{tEshort45}
Selected Ni$_{3}$V energies (meV/atom) from DFT and CE
 relative to $E^{D0_{22}}_{DFT}$, along with those of disordered state, 
its approximation by $SQS_{16}$ \cite{WolvertonSQS16}, and $(001)$ APB.
}
\begin{tabular}{l|r|rr|rrr}
\hline
Structure & $E_{DFT}$ & $E_{CE_2}$ & $\delta E_{CE_2}$ & $E_{CE_3}$ & $\delta E_{CE_3}$ & \\
\hline
$D0_{22}$ $(1 \frac{1}{2} 0) $ & 0.0 & 7.1 & 7.1 & 10.5 & 10.5 & \\
$D0_{23}$ $(1 \frac{1}{4} 0) $ & 25.4 & 31.6 & 6.2 & 35.8 & 10.4 & \\
& 33.7 & 31.6 & -2.1 & 35.8 & 2.1 & \\
$L1_2$ $(1 0 0)$ & 101.2 & 56.1 & -45.1 & 61.1 & -40.1 & \\
& 147.8 & 170.6 & 22.8 & 154.7 & 6.9 & \\
SQS$_{16}$ & 155.4 & 135.8 & -19.7 & 152.0 & -3.5 & \\
\hline
Disordered & n/a & 109.8 &   & 115.1 & & \\
$(001) APB$ & 101.6 & 98.0 & -3.6 & 101.1 & -0.5 & \\
$L1_2-D0_{22}$ & 101.2 & 49.0 & -52.2 & 50.5 & -50.7 & \\
$D0_{23}-D0_{22}$ & 25.4 &24.5 & -0.9 & 25.3 & -0.1 & \\
\hline
\end{tabular}
\end{table}

{\par}{\em Application:}
We now construct and assess the new
canonical CE for Ni$_3$V based on 45 fully-relaxed 
structural DFT energies,
with relative accuracies of  $\sim$$1\,$meV/atom
\footnote{
   We used the Vienna {\it ab initio} simulation package
  \cite{VASP1,VASP3,VASP4} with
   ultra-soft pseudo-potentials \cite{VanderbiltPP}
   from Kresse and Hafner \cite{KresseHafnerPP},
   a plane-wave cutoff of $440\,$eV, and 
   a fine $k$-space mesh to ensure
   forces $< 30\,$meV/{\AA} and relative energy accuracy 
   $\sim 1\,$meV/atom.}.
A selected set of energies is given in Table \ref{tEshort45}.

\begin{figure}[t]
  \includegraphics[width=80mm]{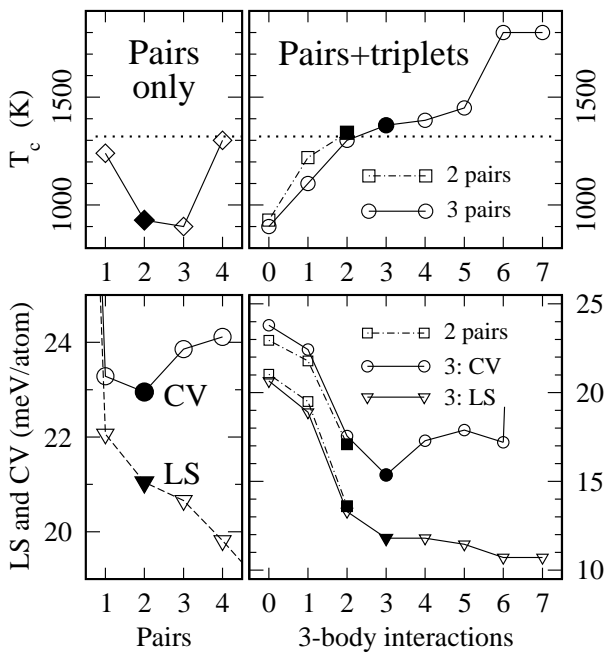}
\setlength{\unitlength}{1mm}
\caption{\label{FigCV2}
(upper) The predicted  and observed (dotted line) $T_c\,$(K),
 and (lower) the CV and LS errors
(meV/atom) vs. ECI truncation.
Results are for CE with pairs-only (left)
and for $2$ and $3$ pairs with a number of 3-body ECI (right).
The minimal CV and LS are given by filled symbols.
}
\end{figure}

{\par}
To examine effects of the new truncation method on prediction, we first
limit the CE basis to pairs only and find that the pairs-only CV is
minimal for 2 (nearest and next-nearest) pair interactions,
see Fig.~\ref{FigCV2}. 
 Within this range, the symmetry-distinct clusters are two pairs, two $3$-bodies, 
 three $4$-bodies, a $5$-body pyramid and a $6$-body octahedron, 
 see Fig.~\ref{FigClusters}. 
The CE with two pair and two triplet interactions (denoted CE$_2$)
with minimal CV of $15.5\,$meV within this range gives $T_c$ at $1335\,$K, 
near the observed $1318\,$K.
Including  tetrahedron does not significantly alter $T_c$, 
as expected for a $4$-body cluster 
at $c=1/4$, as $c^4$$ \ll $$1$ and $V^{(4)}\Phi^{(4)} $$\to$$ 0$.
For optimal truncation, including the most compact 4-body cluster  
is necessary before including any other more extended 
4-body or higher-order clusters.
Within our set of 45 arbitrary structures only one 
($147.8\,$meV/atom in Table \ref{tEshort45})
has contribution from the most compact 4-body cluster, 
so any CE including this cluster has formally infinite (thus not minimal) CV, 
so an optimal CE should contain pairs and triplets only.
Indeed, the optimal CE (denoted CE$_3$), see Fig.~\ref{FigCV2},
includes 3 pair and 3 triplet interactions and yields $T_c$ of $1370\,$K,
again near the observed value and well within the CV error of 
$\pm 15.2\,$meV. Both CE$_2$ and CE$_3$ are examples of the 
localized CE hierarchy allowed by Rules 1--3. 
Notably, we find no failure or ill-description of thermodynamics 
for optimal truncation embodied in the new rules, see Table \ref{tExptCE}.

{\par} Moreover, we predicted from these optimal CE that Ni$_3$V has 
numerous metastable long-period superstructures (LPS) of the type
$\langle 0 \frac{1}{2m} 1 \rangle$ with $m \ge 1$.
The $D0_{22}$ ground state is $m=1$, $D0_{23}$ is $m=2$, and
$L1_2$ is $m=\infty$, see Table \ref{tEshort45}.
We then confirmed by direct DFT calculations that
over 23 metastable structures are between $D0_{22}$ and $L1_2$.
Clearly, structural energy differences then will be sensitive to thermal
antisites or partial-order, {\it i.e.} chemical environments that
distinguish $D0_{22}$ from other low-energy structures.

{\par} The relative energies of $D0_{22}$ and $D0_{23}$, which can be
viewed as $D0_{22}$ with $(001)$ APBs, give an estimate of the
$(001)$ APB energy per site of the antiphase plane:
$ E^{APB} = 4 [ E^{D0_{23}} - E^{D0_{22}} ]$.
In Table \ref{tEshort45}, our calculated $E_{DFT}^{APB}=101.6\,$meV 
and CE$_3$-predicted $E_{CE}^{APB}=101.1\,$meV agree at 
perfect long-range order.
However, binaries with first-order transitions have order parameters $\eta$
(defined in \cite{Johnson2000}) that jump from $0$ to $0.7$--$0.9$ at $T_c$.
For partial order below $T_c$  as in experiment, 
we predict that $E^{APB}_{CE}(\eta )$ are $81$, $65$, and $50\,$meV 
for $\eta$'s of $0.9$, $0.8$, and $0.7$, respectively. 
From superdislocation separation measurements, assessed values are
$52 \pm 20\,$meV at $273\,$K and 
$55 \pm 18\,$meV at $900\,$K \cite{Finel1994}, with roughly constant
 $\eta < 1$ due to lack of kinetics.

%
{\par} The real-space Warren-Cowley SRO parameters $\alpha_{lmn}$ 
were calculated using our CE within Monte Carlo      
at $T$$=$$1.04~T_c$, as in experiment.
The proper way to compare calculated SRO to
experimental data is in~[\onlinecite{Ag2Al}].
Full details of the agreement between calculated and experimental
$\alpha_{lmn}$ will be given elsewhere.
However the energetics associated with SRO given by
$\Delta E_{SRO}^{L1_2 - DO_{22}} \approx k_B T
\left(\alpha^{-1} (1 0 0) - \alpha^{-1} (1 {\frac{1}{2}} 0) \right)/16c(1-c)$
can be directly estimated from the calculated
$\alpha ({\bf k})$ at \{$1 {\frac{1}{2}} 0$\} and \{$1 0 0$\}
{\bf k}-points, as done experimentally \cite{Finel1994}.
We obtain $17\pm15~$meV for CE$_3$ at $1392\,$K, now in agreement with
experiment  \cite{Finel1994}  and 
 coherent-potential approximation (CPA) results  \cite{Johnson2000}, 
see Table \ref{tExptCE}.
Our results confirm the CPA explanation for Ni$_3$V SRO energetics 
and the discrepancy between $T=0\,$K DFT results and measurements 
as arising from the strong dependence of $E^{L1_2}(\eta)$ 
on the state of partial order \cite{Johnson2000}.

%
{\par} Finally, we discuss issues that led to  previous poor 
Ni$_3$V results.
For the range of ECIs included in our {\em truncated} CE basis sets, 
$ \Delta E_{CE}^{L1_2-D0_{22}}
{\equiv} [E_{CE}^{L1_2} - E_{CE}^{D0_{22}}] =
          2 [E_{CE}^{D0_{23}} - E_{CE}^{D0_{22}}] 
{\equiv} E^{APB}/2 $, 
as verified in Table~\ref{tEshort45}.
So for a truncated CE, $L1_2$ can be viewed also as a
$(001)$ APB in $D0_{22}$.
Other LPS, e.g., with $E_{DFT}=33.7\,$meV/atom and
$E^{DO_{23}}_{DFT}=25.4\,$meV/atom in Table~\ref{tEshort45},
also have indistinguishable energies within the truncated CE.
This observation has great import in Ni$_3$V. 
Table~\ref{tEshort45} shows that
$\Delta E^{L1_2 - DO_{22}}_{CE}$ and
$\Delta E^{L1_2 - DO_{22}}_{DFT}$ are not equal!
This implies again that there is a strong configurational
dependence of partially-ordered ${L1_2}$ energy on  $\eta$,
as indeed shown by  CPA calculations \cite{Johnson2000}.
Because $L1_2$ is highly metastable with respect to $D0_{22}$,
a truncated CE will be suspect versus $\eta$ 
(particularly for $\eta \sim 1$) unless all clusters that distinguish 
$L1_2$ from $D0_{22}$, and similar LPS, are included in the basis.
In Refs.~\onlinecite{PRB50p6626,PRB49p16058,PRB52p8813},
$\Delta E^{L1_2 - DO_{22}}_{CE}$ at $0\,$K was forced
to coincide with $\Delta E^{L1_2 - DO_{22}}_{DFT} = 101\,$meV
by overfitting and including certain $3$- and  $4$-body clusters arbitrarily 
that created an invalid CE and hence inaccurate energetics.
Our truncated CE properly describes the observed thermodynamics,
but not high-energy and  ill-described  structures like fully-ordered 
$L1_2$ that are unimportant  for thermodynamics, 
as evidenced in Table~\ref{tEshort45}.
Of course, calculating more DFT structural energies and properly
extending the CE basis to include critical $n$-body clusters
would describe everything more reliably.

{\par} Generally speaking,
neglecting stronger interactions and assigning their weight 
to weaker and less physically important longer-range [or higher-order] 
ones,  {\it i.e.} violating Rule 1 [or Rule 2],
leads to inaccurate predictions of energetics. 
Overfitting (neglecting Rule 3) results in large errors 
in predicted energies,  which were not used in the fit. 
Combined violations can result in dramatic failures:
for example, previous CE for Ni-V \cite{PRB49p16058} 
overfitted energies (violating Rule 3) and included, {\emph e.g.},
a longer-ranged 4-body before the most compact one (violating Rule 1) 
 -- hence the disagreement with experiment and CPA.  
Previous CE results are generally valid if only Rules 1 and 2 were 
obeyed with no large overfitting; in such cases, minimizing the CV 
error leads only to a moderate improvement of accuracy. 
The optimal CE basis truncation
presented has an error that is variational, which is not necessarily 
the case for other basis-set truncation and reduction methods.

{\par}
In summary, the cluster-expansion method is a valuable first-principles 
tool for predicting and interpreting thermodynamic behavior in alloys.
With convergence and reliability in mind, we presented a simple
variational method for the optimal truncation of the cluster-expansion 
basis set.
We presented the method as a set of rules that are computationally
easy to implement.
For a given set of DFT structural energies and no {\em a priori} 
knowledge of which clusters are needed to represent a 
particular alloy well, our truncation method provides
a unique optimal choice of clusters based on their contribution 
to thermodynamics and variational reduction in error, 
avoids choosing clusters by intuition,
and gives reliable thermodynamic predictions.
We exemplified the importance of this new approach in fcc Ni$_3$V
by predicting important new metastable structures and
by showing agreement with order-disorder temperature, 
anti-phase boundary energy, and short-range order energetics,
all quantities missed by previous cluster-expansion applications.
We also elucidated the origin of the  previous failures. 
Without {\em a priori} information, the new cluster-expansion 
strategy allows reliable thermodynamic predictions in alloys.

\begin{acknowledgments}
We acknowledge support from the NSF through a
FRG at the Materials Computation Center (DMR-99-76550) and
an ITR (DMR-0121695), the U.S. Dept. of Energy through
the Frederick Seitz Materials Research Laboratory (DEFG02-91ER45439)
and an IBM SURS grant.
\end{acknowledgments}

\bibliography{Ni3V-CE}
\end{document}